\documentclass[]{aa}  

\usepackage{graphicx}
\usepackage{txfonts}
\usepackage{color}
\usepackage[colorlinks,citecolor=blue]{hyperref}
\usepackage{multirow}  

\begin{document} 

\title{Photoionized emission and absorption features in the high-resolution X-ray spectra of NGC\,3783}

\author{Junjie Mao\inst{\ref{inst_sron},\ref{inst_strath}}
\and M. Mehdipour\inst{\ref{inst_sron}}
\and J. S. Kaastra\inst{\ref{inst_sron},\ref{inst_leiden}}
\and E. Costantini\inst{\ref{inst_sron}}
\and C. Pinto{\inst{\ref{inst_ioa}}}
\and G. Branduardi-Raymont\inst{\ref{inst_mssl}}
\and E. Behar\inst{\ref{inst_tiit}}
\and U. Peretz\inst{\ref{inst_tiit}}
\and S. Bianchi\inst{\ref{inst_ur3}}
\and G. A. Kriss\inst{\ref{inst_stsi}}
\and G. Ponti\inst{\ref{inst_mpe}}
\and B. De Marco\inst{\ref{inst_ncac}}
\and P.-O. Petrucci\inst{\ref{inst_uga}}
\and L. Di Gesu\inst{\ref{inst_ug}}
\and R. Middei\inst{\ref{inst_ur3}}
\and J. Ebrero\inst{\ref{inst_esa}}
\and N. Arav{\inst{\ref{inst_vt}}}
}

\institute{SRON Netherlands Institute for Space Research, Sorbonnelaan 2, 3584 CA Utrecht, the Netherlands\label{inst_sron} \\ \email{J.Mao@sron.nl}
\and Department of Physics, University of Strathclyde, Glasgow, G4 0NG, UK\label{inst_strath}
\and Leiden Observatory, Leiden University, Niels Bohrweg 2, 2300 RA Leiden, the Netherlands\label{inst_leiden}
\and Institute of Astronomy, University of Cambridge, Madingley Rd, Cambridge, CB3 0HA, UK\label{inst_ioa}
\and Mullard Space Science Laboratory, University College London, Holmbury St. Mary, Dorking, Surrey, RH5 6NT, UK\label{inst_mssl}
\and Department of Physics, Technion-Israel Institute of Technology, 32000 Haifa, Israel\label{inst_tiit}
\and Dipartimento di Matematica e Fisica, Universit\`{a} degli Studi Roma Tre, via della Vasca Navale 84, 00146 Roma, Italy\label{inst_ur3}
\and Space Telescope Science Institute, 3700 San Martin Drive, Baltimore, MD 21218, USA\label{inst_stsi}
\and Max Planck Institute fur Extraterrestriche Physik, 85748, Garching, Germany\label{inst_mpe}
\and Nicolaus Copernicus Astronomical Center, Polish Academy of Sciences, Bartycka 18, PL-00-716, Warsaw, Poland\label{inst_ncac}
\and Univ. Grenoble Alpes, CNRS, IPAG, 38000 Grenoble, France\label{inst_uga}
\and Department of Astronomy, University of Geneva, 16 Ch. d'~Ecogia, 1290 Versoix, Switzerland\label{inst_ug}
\and European Space Astronomy Centre, P.O. Box 78, E-28691 Villanueva de la Ca\~{n}ada, Madrid, Spain\label{inst_esa}
\and Department of Physics, Virginia Tech, Blacksburg, VA 24061, USA\label{inst_vt}
}

\date{Received date / Accepted date}

 
\abstract
{Our \textit{Swift} monitoring program triggered two joint \textit{XMM-Newton}, \textit{NuSTAR} and HST observations on 11 and 21 December 2016 targeting NGC\,3783, as its soft X-ray continuum was heavily obscured. Consequently, emission features, including the \ion{O}{VII} radiative recombination continuum, stand out above the diminished continuum. We focus on the photoionized emission features in the December 2016 RGS spectra, and compare them to the time-averaged RGS spectrum obtained in 2000--2001 when the continuum was unobscured. A two-phase photoionized plasma is required to account for the narrow emission features. These narrow emission features are weakly varying between 2000--2001 and December 2016. We also find a statistically significant broad emission component in the time-averaged RGS spectrum in 2000--2001. This broad emission component is significantly weaker in December 2016, suggesting that the obscurer is farther away than the X-ray broad-line region. In addition, by analyzing the archival high-resolution X-ray spectra, we find that nine photoionized absorption components with different ionization parameters and kinematics are required for the warm absorber in X-rays. }


\titlerunning{X-ray photoionized broad and narrow emission features in NGC\,3783}
\authorrunning{Mao et al.}

\maketitle

\section{Introduction}
\label{sct:intro}
The optical spectra of Seyfert 1 galaxies often show broad emission lines with a velocity broadening of a few $10^3~{\rm km~s^{-1}}$ and narrow emission lines with a velocity broadening about a few $10^2~{\rm km~s^{-1}}$ \citep[e.g.,][]{bla90}. In the soft X-ray band of Seyfert 1 galaxies, broad and narrow emission lines are also observed \citep[e.g.,][]{kaa00,ogl04,ste09,lon10,cos07,cos16}. Although in some Seyfert 2 galaxies the spatial extent of the optical and X-ray narrow emission line regions is remarkably similar \citep{bia06}, the relation of narrow emission lines in the optical and X-ray bands are not fully understood. Similarly, the connection of optical to X-ray broad emission lines is also poorly understood. 

Optical broad emission lines are known to vary, as they are dependent on the luminosity of the nucleus \citep[][and references therein]{ben13}. Optical narrow emission lines (e.g. [\ion{O}{III}]) are historically thought to be constant in flux and used to calibrate spectroscopic monitoring data \citep{pet13}. However, there is also a growing number of studies \citep{det09, pet13, den14, lan15} suggesting that the optical and/or X-ray narrow emission lines in active galactic nuclei \citep[AGN,][]{pet97,kro99} are variable over long timescales (at least a few years). This is not totally unexpected considering the photoionization origin of these emission lines, the variable ionizing source and the distance of the narrow-line region \citep[at least a few parsecs,][]{ben06a,ben06b}. 

NGC\,3783 is a nearby \citep[$z=0.009730$,][]{the98} Seyfert 1 galaxy that has been extensively studied for the past few decades in the infrared, optical, UV and X-ray wavelength ranges \citep[e.g.,][]{kas00,kra01,onk02,ram05,che05,hon13,goo16,fuk18}. NGC\,3783 has a supermassive black hole with $M_{\rm BH}=(3.0\pm0.5)\times10^{7}~{\rm M_{\odot}}$ \citep{pet04}. The radius of the optical broad-line region in NGC\,3783 ranges from 0.0012~pc (or 1.4 light-days for \ion{He}{II}) to 0.0086~pc \citep[or 10.2 light-days for H$\beta$,][]{pet04}. 

On 11 and 21 December 2016, our \textit{Swift} monitoring program\footnote{The X-ray hardness variability of eight type-I AGN was monitored with \textit{Swift} in 2016 in order to find intense obscuration events and thereby study them with joint Hubble Space telescope, \textit{XMM-Newton}, and \textit{NuSTAR} observations \citep{meh17}.} triggered two joint \textit{XMM-Newton}, \textit{NuSTAR} and Hubble Space Telescope (HST) observations targeting NGC\,3783 \citep{meh17}. This is because the soft X-ray continuum of NGC\,3783 was heavily obscured \citep[See Fig.~2 in][]{meh17}. This is similar to the obscuration events discovered in NGC\,5548 \citep{kaa14} and NGC\,985 \citep{ebr16}, where strong absorption of the X-ray continuum and outflowing UV broad absorption lines are observed simultaneously. The obscurer in NGC\,3783 is found to partially cover the central source with a column density on the order of $10^{27}~{\rm m^{-2}}$ \citep{meh17}. The obscurer is outflowing with a range of velocities up to $6000~{\rm km~s^{-1}}$ and it is probably a disk wind at the outer broad-line region of the AGN \citep{meh17}. Moreover, based on X-ray data accumulated in the past two decades, \citet{kaa18} suggest that obscuration events with $N_{\rm H}\gtrsim10^{26}~{\rm m^{-2}}$ are a frequent phenomenon in NGC\,3783. Based on a detailed UV spectral analysis, \citet{kri18} suggest that a collapse of the broad-line region has led to the outburst and triggered the obscuring event.  Several other examples of transient X-ray obscuration without simultaneous UV observations have also been observed \citep[e.g.,][]{riv15,beu17,par17,tur18}.

Here we present a study of photoionized emission features in the soft X-ray band of NGC\,3783 from our recent observations in December 2016 and archival observations in 2000--2001. This paper is structured as follows: In Section~\ref{sct:obs}, we list all the observations used. Then we detail the spectral analysis in Section~\ref{sct:spec}, including the data treatment and the model description, which has been briefly mentioned in \citet{meh17}. We present the results of our spectral analysis in Section~\ref{sct:res}, as well as discussions of physical implications. Conclusions are available in Section~\ref{sct:con}.

\section{Observations and data reduction}
\label{sct:obs}
In 2000--2001 absorption features in NGC\,3783 caused by the warm absorber were clearly detectable, as well as the \ion{O}{VIII} and \ion{C}{VI} Ly$\alpha$ emission lines and the He-like triplets of \ion{O}{VII} \citep{kas00,blu02}. However, in December 2016, the soft X-ray continuum was heavily absorbed by the obscurer so that narrow absorption features caused by the warm absorber were hardly visible. Due to this diminished continuum, narrow emission features are better visible in December 2016. This includes \ion{O}{VII} narrow radiative recombination continuum (RRC), which is a characteristic emission feature of a photoionized plasma \citep{lie96}. 

The spectral analysis of such complicated spectra depends crucially on disentangling the true continuum from the effects of the obscuration and absorption features, which may include many lines and edges that are unresolved and overlapping, as well as emission features. That is to say, this requires a model capable of fitting essentially the overall continuum as well as all the absorption and emission features, including those that cannot be clearly resolved. 

Therefore, as described in \citet{meh17}, we construct a time-averaged spectrum in 2000 and 2001 to constrain the intrinsic spectral energy distribution (SED) and emission features in the unobscured state. Archival data from the optical monitor \citep[OM,][]{mas01}, Reflection Grating Spectrometer \citep[RGS,][]{dhe01}, and European Photon Imaging Camera-pn \citep[EPIC-pn,][]{str01} from \textit{XMM-Newton} are used. We also use archival \textit{Chandra} High-Energy Transmission Grating Spectrometer \citep[HETGS,][]{can05} data in 2000, 2001 and 2013 to better constrain the photoionized absorption features, which are rather stable over the 12-year period \citep{sco14}. The time-averaged spectrum has a total exposure of 1.37~Ms, with 1.05~Ms for \textit{Chandra}/HETGS and 324 ks for \textit{XMM-Newton}/RGS , respectively. 

The optical to hard X-ray data in December 2016 are used for determining the intrinsic SED, obscuration effect, and photoionized emission features (present work) in the obscured state. The two data sets in December 2016 are fitted independently as the continua are different \citep[e.g. Figure~2 in][]{meh17}. 

All the data used for the present work are listed in Table~\ref{tbl:obs_log}. The exposures of the obscured \textit{XMM-Newton}/RGS spectra are 110 ks and 56 ks, respectively. In addition, we have UV flux from HST Cosmic Origins Spectrograph \citep[COS,][]{gre12}. A detailed description of the data reduction can be found in Appendix~A of \citet[][for NGC\,5548]{meh15}, which also applies to the NGC\,3783 data used here.

\begin{table*}
\caption{List of NGC\,3783 data used for the present spectral analysis.}
\label{tbl:obs_log}
\centering
\begin{tabular}{cccccccccccccc}
\hline\hline
\noalign{\smallskip}
Observatory & Inst. & ObsID & Start date & Length  \\
\noalign{\smallskip} 
\hline
\noalign{\smallskip} 
\textit{Chandra} & ACIS/HETG & 373 & 2000-01-20 & 56~ks \\
\noalign{\smallskip} 
\textit{XMM-Newton} & OM, RGS, pn & 0112210101 & 2000-12-28 & 40~ks \\
\noalign{\smallskip}
\textit{Chandra} & ACIS/HETG & 2090 & 2001-02-24 & 166~ks \\
\noalign{\smallskip}
\textit{Chandra} & ACIS/HETG & 2091 & 2001-02-27 & 169~ks \\
\noalign{\smallskip}
\textit{Chandra} & ACIS/HETG & 2092 & 2001-03-10 & 165~ks \\
\noalign{\smallskip}
\textit{Chandra} & ACIS/HETG & 2093 & 2001-03-31 & 166~ks \\
\noalign{\smallskip}
\textit{Chandra} & ACIS/HETG & 2094 & 2001-06-26 & 166~ks \\
\noalign{\smallskip}
\textit{XMM-Newton} & OM, RGS, pn & 0112210201/0401 & 2001-12-17 & 142~ks \\
\noalign{\smallskip}
\textit{XMM-Newton} & OM, RGS, pn & 0112210501/0601 & 2001-12-19 & 142~ks \\
\noalign{\smallskip}
\textit{Chandra} & ACIS/HETG & 14991 & 2013-03-25 & 59~ks \\
\noalign{\smallskip}
\textit{Chandra} & ACIS/HETG & 15626 & 2013-03-27 & 102~ks \\
\noalign{\smallskip}
\hline
\noalign{\smallskip}
\textit{HST} & COS & LD3E03 & 2016-12-12 & 2 orbits \\
\noalign{\smallskip}
\textit{XMM-Newton} & OM, RGS, pn & 0780860901 & 2016-12-11 & 110~ks \\
\noalign{\smallskip} 
\textit{NuSTAR} & FPMA, FPMB & 80202006002 & 2016-12-11 & 56~ks \\
\noalign{\smallskip} 
\hline
\noalign{\smallskip}
\textit{HST} & COS & LD3E04 & 2016-12-21 & 2 orbits \\
\noalign{\smallskip}
\textit{XMM-Newton} & OM, RGS, pn & 0780861001 & 2016-12-21 & 56~ks \\
\noalign{\smallskip} 
\textit{NuSTAR} & FPMA, FPMB & 80202006004 & 2016-12-21 & 46~ks \\
\noalign{\smallskip} 
\hline
\end{tabular}
\end{table*}

\section{Spectral analysis}
\label{sct:spec}
The spectral analysis package SPEX \citep{kaa96} v3.04 is used, incorporating state-of-the-art atomic data, including radiative recombination \citep{bad06,mao16} and electron energy loss due to radiative recombination \citep{mao17a}. We use $C$-statistics following \citet{kaa17a} throughout this work. Statistical errors are quoted at 68.3\% ($1\sigma$) confidence level ($\Delta C$ = 1.0) unless indicated otherwise. X-ray spectra are optimally binned according to \citet{kaa16}. Spectra shown in this paper are background subtracted and displayed in the observed frame.

\subsection{Optical to X-ray spectra construction}
\label{sct:spec_data}
The unobscured (2000--2001) and obscured (December 2016) spectra are constructed with all available data in each epoch as follows:
\begin{enumerate}
    \item OM flux at V, B, U, UVW1, UVM2, and UVW2 bands;
    \item COS flux at 1139~\AA, 1339~\AA, 1477~\AA, and 1794~\AA, free from emission and absorption features;
    \item first-order RGS data (RGS1 and RGS2 combined) in the $7-37$~\AA~wavelength range, in order to better constrain the emission and absorption features especially for $\lambda\gtrsim25$~\AA;
    \item first-order MEG (Medium Energy Grating) data (positive and negative orders combined) in the $6-17$~\AA~wavelength range, which have better energy resolution compared to the RGS data at the same wavelength range. Note that MEG data are broken into several segments, with the local continuum of each segment re-scaled by $\lesssim \pm15$~\% to match the RGS continuum at the same wavelength range. The scaling factor of each segment is 1.158 for 6--7~\AA, 1.120 for 7--8~\AA, 1.003 for 8--9~\AA, 1.029 for 9--11~\AA, 0.889 for 11--14~\AA, and 0.850 for 14--17~\AA, respectively. This is to account for observations taken at different epochs, as well as the cross calibration between RGS and MEG; 
    \item first-order HEG (High Energy Grating) data (positive and negative orders combined) in the $1.7-3$~\AA~wavelength range, which has better energy resolution compared to the pn data in the same wavelength range. The HEG data are re-scaled by a factor of 1.218;
    \item EPIC-pn data in the $1.5-10$~keV energy range (i.e. the 1.24--8~\AA~wavelength range). EPIC-pn data are re-scaled by 1.038 (2000--2001);
    \item \textit{NuSTAR} data in the $10-78$~keV energy range. \textit{NuSTAR} data are re-scaled by 1.013 (11 December 2016) and 1.027 (21 December 2016), respectively. 
\end{enumerate}

The above optical to X-ray data are fitted simultaneously, which allows us to constrain the broadband (optical, UV and X-ray) continuum and the X-ray obscuration, absorption and emission features at the same time. Note that the COS grating spectra are analyzed in a separate paper \citep{kri18}. 

We used limited but adequate overlapping data to determine the cross calibration among instruments. This is due to the fact that the cross calibration is not a strict constant across the wavelength/energy range. Introducing more overlapping data would introduce more systematic uncertainties, leading to poorer statistics in the end. Nonetheless, we should emphasize that we use the best instrument for each wavelength/energy range.

\subsection{Description of model components}
\label{sct:spec_model}
In order to interpret the continuum, absorption and emission parts of the spectrum at the same time, following Section 2.2 of \citet{kaa18}, we take model components listed below into account:
\begin{enumerate}
    \item The intrinsic broadband spectral energy distribution (SED) of the AGN.
    \item The obscuration effect caused by the obscurer and the high-ionization component in the obscured state.
    \item Absorption features caused by the warm absorber. 
    \item Broad and narrow emission features caused by the X-ray photoionized emitter.
    \item Broad and narrow emission lines in the optical/UV. 
    \item The host galaxy continuum emission in the optical/UV.
    \item The Galactic extinction in the optical/UV and absorption in X-rays. 
\end{enumerate}
While model components 1, 3, and 4 are described in the following subsections, we refer readers to \citet{meh17} and \citet{kaa18} for details of model components 2, 5, 6 and 7. In Appendix~\ref{sct:com_rel} we discuss the spatial relationships among these components and describe in detail how we combine these components in SPEX. The protosolar abundances of \citet{lod09} are used for all plasma models. 

\subsubsection{Intrinsic broadband SED of the AGN}
A photoionization continuum is required for photoionization modeling of the obscurer, warm absorber, and X-ray photoionized emitter. Following previous analysis of NGC\,5548 \citep{meh15}, we fit the optical to X-ray data of NGC\,3783 using a model consisting of a Comptonized disk component \citep[COMT,][]{tit94} for optical to soft X-rays, a power-law component (POW), and a neutral reflection component \citep[REFL,][]{mag95,zyc99} for hard X-rays. An exponential cut-off is applied to the high- and low-energy end of the power-law component (Appendix~\ref{sct:com_rel}). The intrinsic optical to X-ray continuum and all the obscuration, absorption, emission, and extinction effects are fitted simultaneously. Therefore, the fit iterates many times to get the best-fit intrinsic broadband SED of the AGN (results in Section~\ref{sct:res_cont}). 

\subsubsection{Photoionization continuum}
A photoionization continuum is required for the photoionization modeling and it can be different from the AGN SED mentioned above. The photoionization continuum received by the obscurer is indeed simply the contemporary AGN SED (Table~\ref{tbl:sed_par}). 

For the warm absorber, in 2000--2001, when NGC\,3783 is not obscured, its photoionization continuum is also the contemporary AGN SED. However in December 2016 the photoionization continuum for the warm absorber is the contemporary AGN SED with the obscuration effect taken into account. 

The photoionization continuum received by the X-ray narrow-line region is assumed to be the 2000--2001 (time-averaged) AGN SED for all epochs (2000--2001 and December 2016). This is because if the photoionized plasma is too far away from the nucleus and/or the density of the plasma is too low, the plasma is in a quasi-steady state with its ionization balance varying slightly around the mean value corresponding to the mean ionizing flux level over time \citep{nic99,kaa12,sil16}. Note that our assumption implies that the obscurer does not subtend a large solid angle so that it barely screens photons from the nucleus to the X-ray narrow emission component or the screened photons have not arrived the X-ray narrow emission component yet. 

\subsubsection{Warm absorber}
We use the photoionization plasma model PION in SPEX to account for the absorption features produced by the warm absorber. PION instantaneously determines the thermal equilibrium, ionization balance and level populations of all ions from the input SED and produces an absorption (or emission) model spectrum. As described in Section~\ref{sct:res_wa}, we require nine PION absorption components. For the time-averaged unobscured spectrum (2000--2013), we allow hydrogen column densities ($N_{\rm H}$), ionization parameters ($\log \xi$), and outflow velocities ($v_{\rm out}$) free to vary. After a few trials, microscopic turbulence velocities\footnote{The Doppler parameter $b=(v_{\rm th}^2 + 2 v_{\rm mic}^2)^{1/2}$, where $v_{\rm th}=\sqrt{2kT/m}$ is the thermal broadening velocity, $T$ the temperature, and $m$ the atomic mass. The full width at half maximum ${\rm FWHM} = 2\sqrt{\ln~2}~b$. In the PION model in SPEX, $v_{\rm mic}$ is the ``v" parameter. } ($v_{\rm mic}$) are coupled for components 1--2, 3--7, and 8--9 to reduce unnecessary free parameters. For the obscured spectra (December 2016), all the above parameters are fixed to values obtained in the unobscured spectrum, except the ionization parameters, which are assumed to be proportional to the $1-10^3$~Ryd ionizing luminosity. That is to say, the hydrogen number density times distance squared ($n_{\rm H} r^2$) of the warm absorber in the obscured state is assumed to be the same as in the unobscured state. In both unobscured and obscured states, each PION absorption component fully covers the line of sight ($C_{\rm abs} = 1$, frozen) but has negligible extent ($C_{\rm em}=\Omega/4\pi=0$, frozen) with respect to the nucleus. 

\subsubsection{X-ray photoionized emitter}
Similar to the previous analysis on NGC\,5548 \citep{mao18a}, the narrow emission features in NGC\,3783 can be reasonably fitted with two PION components (details in Section~\ref{sct:res_xpe}). Whether the soft X-ray continuum is obscured or not, four free parameters of each PION emission component are allowed to vary, namely $N_{\rm H}$, $\log \xi$, $v_{\rm mic}$ and $C_{\rm em}$. Since the emission lines are consistent with not outflowing, as shown by previous studies \citep{kas02,beh03}, the outflow velocity ($v_{\rm out}$) is fixed to zero for each PION emission component. In addition, for each PION emission component, its absorption covering factor $C_{\rm abs}$ is fixed to zero. 

Broad emission features are modeled with a third PION component. For the time-averaged unobscured spectrum (2000--2001), where the \textit{XMM-Newton} and \textit{Chandra} grating spectra are fitted simultaneously, three free parameters of this PION emission component are allowed to vary, including $N_{\rm H}$, $\log \xi$, and $C_{\rm em}$. For the obscured spectra (December 2016), these parameters are fixed to values obtained in the unobscured spectrum, except the ionization parameters, which are assumed to be proportional to the $1-10^3$~Ryd ionizing luminosity (Table~\ref{tbl:sed_par}). 

For simplicity, all three emission components are assumed to be free of further absorption by the warm absorber (see discussion in Section~\ref{sct:sum}). 

\section{Results and discussions}
\label{sct:res}
\subsection{Intrinsic broadband SED of the AGN}
\label{sct:res_cont}
\begin{figure}
\centering
\includegraphics[width=\hsize, trim={0.0cm, 0.5cm, 0.5cm, 0.5cm}, clip]{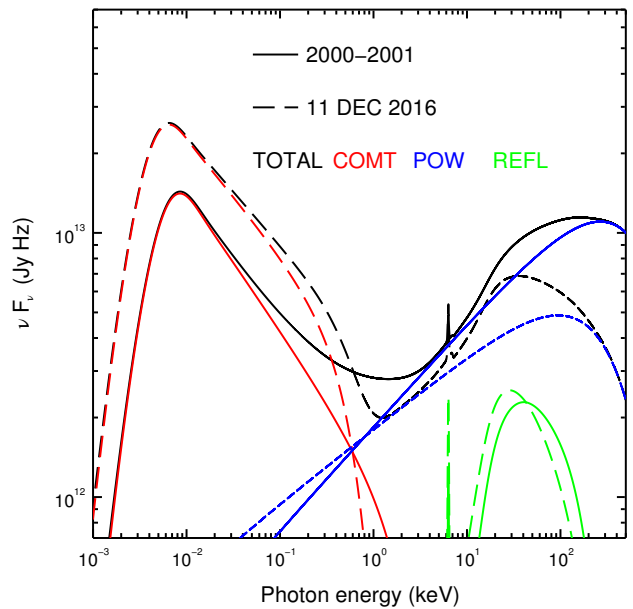}
\caption{The intrinsic spectral energy distribution of NGC\,3783 for the time-averaged data in 2000--2001 (solid lines) and 11 December 2016 data (dashed lines). Contributions from individual components are shown in red for the Comptonized disk component (COMT), blue for the power-law component (POW) and green for the reflection component (REFL).}
\label{fig:sed_plot}
\end{figure}

The time-averaged intrinsic SED of NGC\,3783 in 2000-2013 is shown in Figure~\ref{fig:sed_plot}. The intrinsic SED on 11 December 2016 is also shown for comparison, which is similar to that on 21 December 2016. Fig.~6 of \citet{meh17} presents a version with the intrinsic SEDs overplotted with the observational data. The corresponding best-fit continuum parameters are listed in Table~\ref{tbl:sed_par}. Figure~\ref{fig:sed_plot} shows the comparison of the intrinsic and obscured SED of NGC\,3783 on 11 December 2016.

\begin{figure}
\centering
\includegraphics[width=\hsize, trim={0.0cm, 0.5cm, 0.5cm, 0.5cm}, clip]{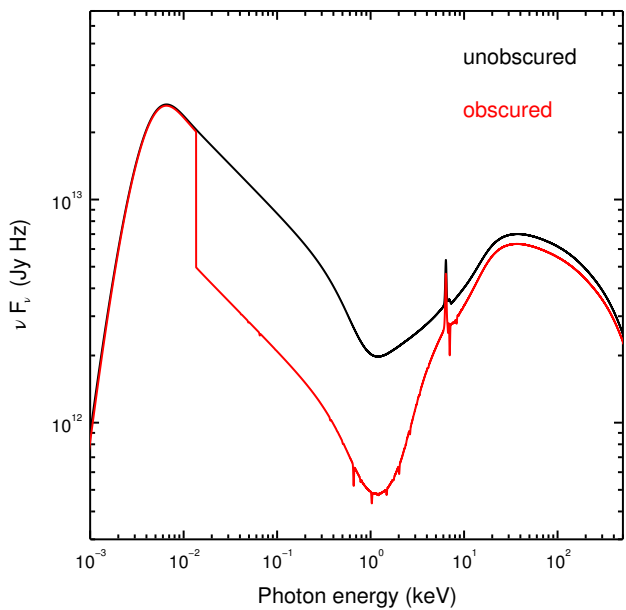}
\caption{The intrinsic/unobscured (black) and obscured (red) spectral energy distribution of NGC\,3783 on 11 December 2016 (solid lines).}
\label{fig:osed_plot}
\end{figure}

\begin{table*}
\caption{Best-fit parameters of the intrinsic broadband SED of NGC\,3783 in 2000-2001 and December 2016.}
\label{tbl:sed_par}
\centering
\begin{tabular}{lccccccccccccc}
\hline\hline
\noalign{\smallskip}
Date & 2000-2001 & 11 DEC 2016 & 21 DEC 2016 \\
\noalign{\smallskip}
\hline
\noalign{\smallskip}
\multicolumn{4}{c}{COMT} \\
\noalign{\smallskip}
Norm & $0.95\pm0.05$ & $9.7\pm0.3$ & $10.3\pm0.4$ \\
\noalign{\smallskip}
$T_{\rm seed}$ & $1.47\pm0.05$ & $1.08\pm0.03$ & $1.08\pm0.02$ \\
\noalign{\smallskip}
$T_{\rm c}$ & $0.54\pm0.02$ & $0.13\pm0.02$ & $0.14\pm0.03$ \\
\noalign{\smallskip}
$\tau$ & $9.9\pm0.2$ & $22.0$ (f) & $22.0$ (f) \\
\noalign{\smallskip}
\multicolumn{4}{c}{POW} \\
\noalign{\smallskip}
Norm & $2.40\pm0.03$ & $2.42\pm0.16$ & $3.03\pm0.19$ \\
\noalign{\smallskip}
$\Gamma$ & $1.61\pm0.01$ & $1.73\pm0.02$ & $1.75\pm0.02$ \\
\noalign{\smallskip}
$C$-stat / $C$-expt. & 6092 / 2563 & 2335 / 1552 & 2340 / 1555 \\
\noalign{\smallskip}
$L_{\rm 0.3-2~keV}$ & 1.20 & 1.28 & 1.52 \\
\noalign{\smallskip}
$L_{\rm 2-10~keV}$ & 1.16 & 0.94 & 1.11 \\
\noalign{\smallskip}
$L_{\rm 0.001-10~keV}$ & 10.6 & 18.8 & 21.1 \\
\noalign{\smallskip}
$L_{\rm 1-1000~Ryd}$ & 7.01 & 10.2 & 11.9 \\
\noalign{\smallskip}
\hline
\end{tabular}
\tablefoot{Parameters followed by (f) are fixed in the fit. The temperature of seed photons is in units of eV, the warm corona temperature in keV, the normalization of the warm comptonization (COMT) component in $10^{55}~{\rm ph~s^{-1}~keV^{-1}}$, and the normalization of the power-law (POW) component in $10^{51}~{\rm ph~s^{-1}~keV^{-1}}$ at 1~keV. The $C$-stat refer to the final best-fit, where all obscuration, absorption, emission and extinction effects are taken into account. The luminosities of COMT plus POW in different energy ranges are in units of $10^{36}~{\rm W}$ with uncertainties about $3-5$\%. All quoted errors refer to the statistical uncertainties at the 68.3~\% confidence level.}
\end{table*}

The two observations in December 2016 have similar intrinsic SEDs. However, for the time-averaged SED in 2000--2013, both the Comptonized disk component and the power-law component differ significantly from December 2016.  

We should emphasize that we apply a global SED model to fit the optical to hard X-ray data, thus, the continuum model in the present work differs from previous studies. \citet{kas01} and \citet{sco14} fitted the 1--26~\AA\ continuum with ``line-free zones" (LFZs). However, as pointed out by \citet{sco14}, even when LFZs are entirely line-free, absorption edges still contribute to curvature in the spectrum. The $0.5-10$~keV continuum of NGC\,3783 has also been fitted by a power law with $\Gamma\in(1.5,~1.83)$ \citep{blu02,dro02}, sometimes including an extra thermal component with $kT\sim0.1$~keV \citep{kro03}. \citet{fu17} constructed a SED with typical AGN SED \citep{elv94} but scaled with wavelength ($\propto\lambda^{-0.27}$) to match the intrinsic UV ($1135-1795$~\AA) and X-ray ($2-11$~\AA) continuum. 

\subsection{Warm absorber}
\label{sct:res_wa}
\begin{figure*}
\centering
\includegraphics[width=\hsize, trim={0.5cm 2.0cm 0.5cm 0.5cm}, clip]{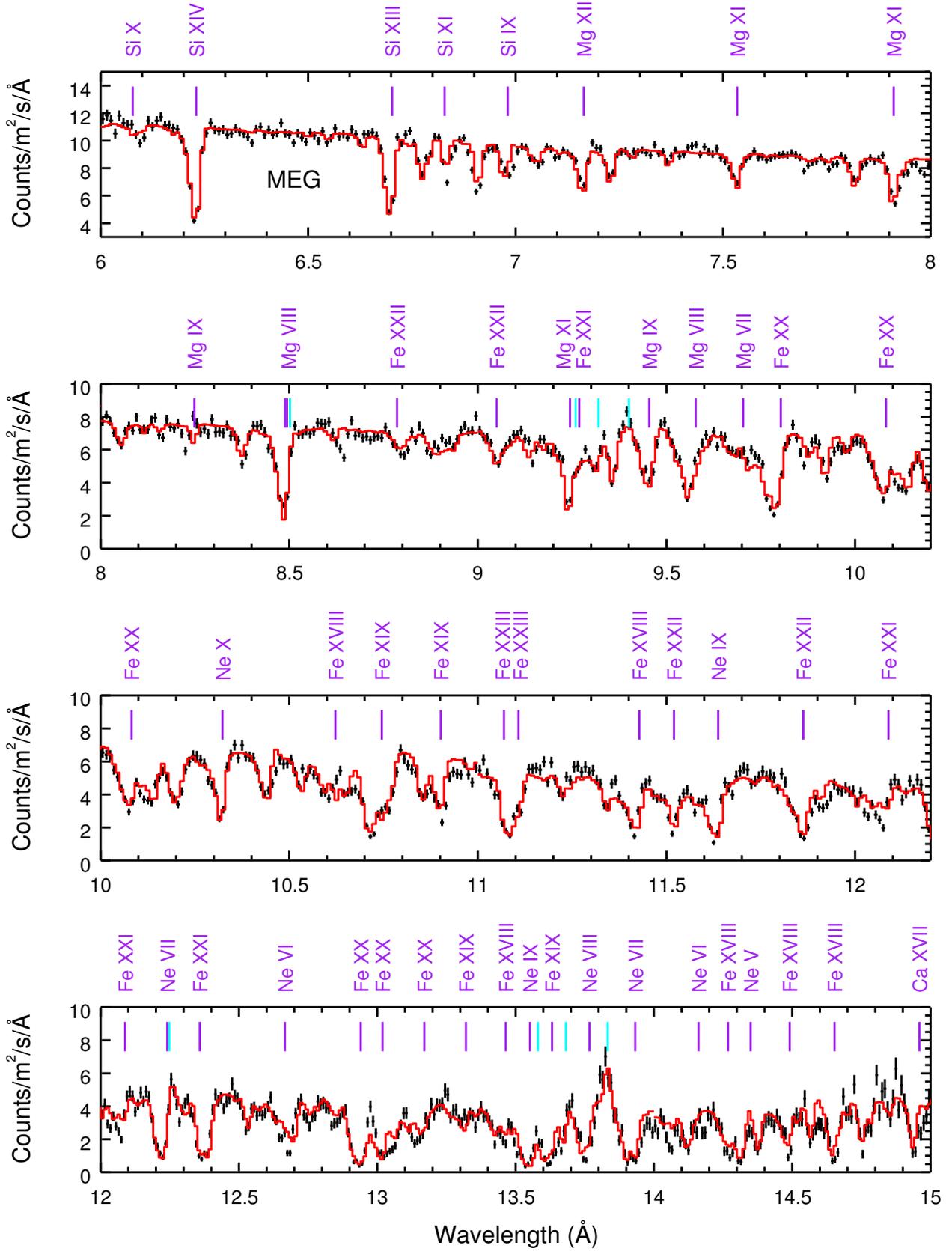}
\caption{The best-fit to the time-averaged MEG spectrum (in the observed frame) of NGC\,3783 observed in 2000, 2001 and 2013. Most prominent absorption and emission features are labeled. The solid vertical lines in purple (blue) indicate the photoionized absorption (emission) features.}
\label{fig:spec_meg_plot}
\end{figure*}

\begin{figure*}
\centering
\includegraphics[width=\hsize, trim={0.5cm 2.0cm 0.5cm 0.5cm}, clip]{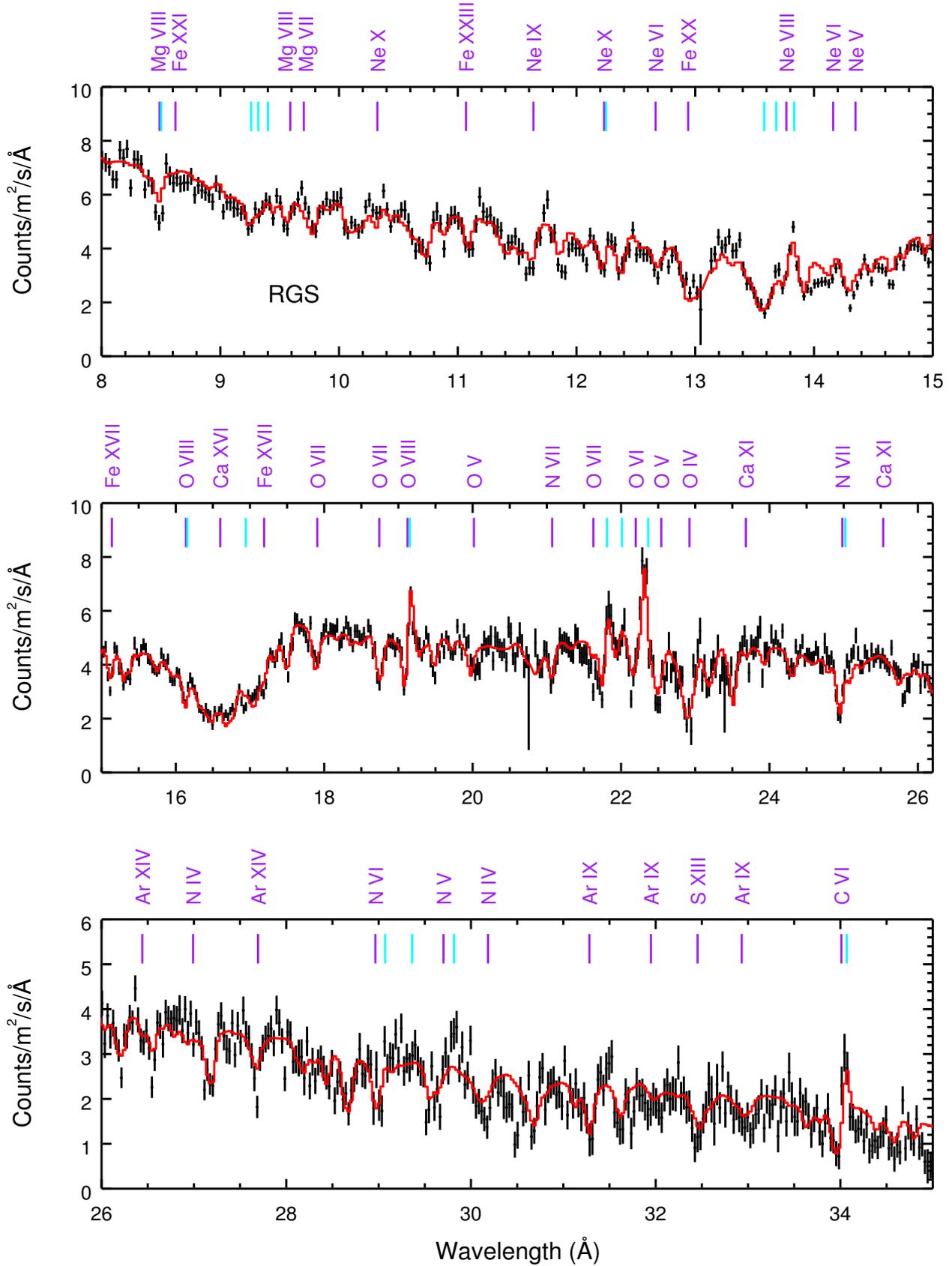}
\caption{Similar to Figure~\ref{fig:spec_meg_plot} but for the time-averaged RGS spectrum (in the observed frame) of NGC\,3783 observed in 2000 and 2001.}
\label{fig:spec_rgs_plot}
\end{figure*}

The best-fit $C$-stat to expected $C$-stat ratio for the time-averaged spectrum in 2000--2013 is $6092/2563\gtrsim2$ (with 2505 degrees of freedom), which is not statistically acceptable. This is due to the very high photon statistics compared with systematic uncertainties in the instrumental response, the imperfect cross calibration of different instruments, and our model is still rather simple compared to the reality. But the result is still useful to assess whether the observed continuum and all absorption (and emission) features are reasonably accounted for. The best-fit time-averaged (2000, 2001 and 2013) MEG spectrum (6--15~\AA) is shown in details in Figure~\ref{fig:spec_meg_plot}. The best-fit time-averaged (2000--2001) RGS spectrum (8--35~\AA) is fitted simultaneously but shown separately in Figure~\ref{fig:spec_rgs_plot} for clarity. 

\begin{figure*}
\centering
\includegraphics[width=\hsize, trim={0.5cm 2.0cm 0.5cm 0.5cm}, clip]{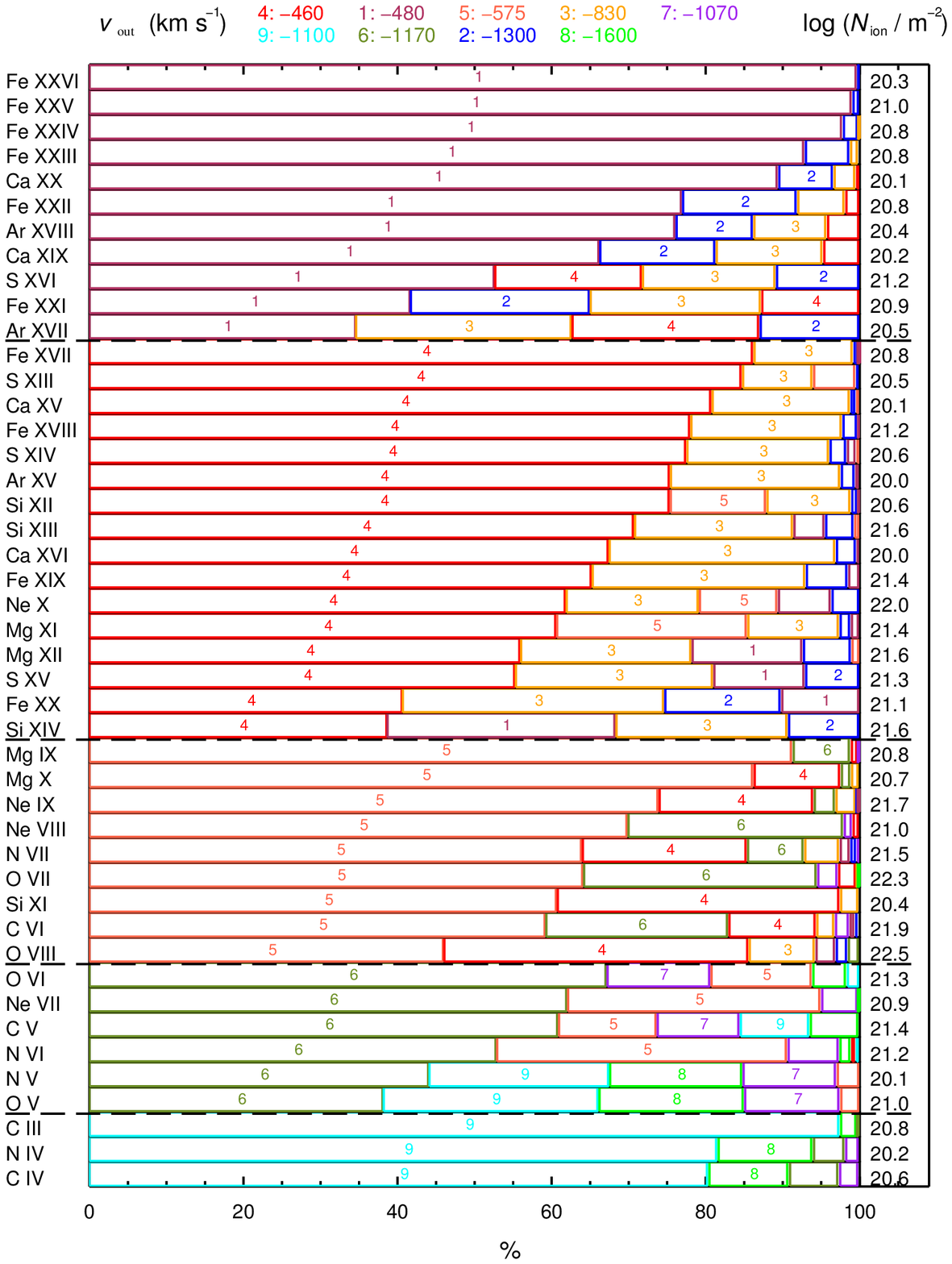}
\caption{Ionic column density (units: ${\rm m^{-2}}$, in log-scale listed to the right) with contributions from individual warm absorber components (in percentage). Ions with $N_{\rm ion}\gtrsim 10^{20}~{\rm m^{-2}}$ are listed to the left. The warm absorber components are color coded with outflow velocities (${v_{\rm out}}$) listed to the top. The ionization parameters of the warm absorber components increase from component 9 to component 1.}
\label{fig:colden_plot}
\end{figure*}

There are in total nine PION absorption components for the warm absorber, with components 1--4 dominating highly ionized ions with large atomic number (e.g. \ion{Fe}{XXVI}, \ion{Fe}{XVII}, and \ion{Si}{XIV}), components 5 and 6 mainly accounting for moderately ionized ions with intermediate atomic number (e.g. \ion{Mg}{IX}, \ion{O}{VIII} and \ion{O}{V}), and components 7--9 for lowly ionized ions with intermediate atomic number (e.g. \ion{N}{IV} and \ion{C}{IV}). Contributions from individual warm absorber components (in percentage) to ions with column density $N_{\rm ion}\gtrsim10^{20}~{\rm m^{-2}}$ are shown in Figure~\ref{fig:colden_plot}. Components 7--9 are required in our model because we include the RGS spectrum above $25$~\AA, while previous analyses simply focus on spectra at shorter wavelength range with $\lambda\lesssim25~\AA$ \citep[][for RGS and MEG, respectively]{beh03,kro03}. Admittedly, the signal-to-noise ratio decreases toward the longer wavelength, thus, the best-fit parameters of components 7--9 are less well constrained. All the best-fit parameters for the warm absorber components of NGC\,3783 in the 2000--2013 time-averaged spectra are tabulated in Table~\ref{tbl:wa_par}. 

\begin{table*}
\caption{Best-fit parameters of the warm absorber components in the time-averaged spectra of NGC\,3783 in 2000--2013.}
\label{tbl:wa_par}
\centering
\begin{tabular}{cccccccccccccccc}
\hline\hline
\noalign{\smallskip}
Comp. & $N_{\rm H}$ & $\log_{10}(\xi)$ & $v_{\rm mic}$ & $v_{\rm out}$ \\
& $10^{25}~{\rm m^{-2}}$ & $10^{-9}~{\rm W~m}$ & ${\rm km~s^{-1}}$ & ${\rm km~s^{-1}}$ \\
\noalign{\smallskip} 
\hline
\noalign{\smallskip} 
1 & $11.1\pm0.8$ & $3.02\pm0.01$ & $120\pm10$ & $-480\pm10$ \\
\noalign{\smallskip} 
2 & $2.1\pm0.2$ & $2.74\pm0.03$ & 120 (c) & $-1300\pm25$ \\
\noalign{\smallskip} 
3 & $6.1\pm0.6$ & $2.55\pm0.02$ & $46\pm2$ & $-830\pm15$ \\
\noalign{\smallskip} 
4 & $12.4\pm0.5$ & $2.40\pm0.01$ & 46 (c) & $-460\pm10$ \\
\noalign{\smallskip} 
5 & $5.0\pm0.2$ & $1.65\pm0.01$ & 46 (c) & $-575\pm10$ \\
\noalign{\smallskip} 
6 & $1.2\pm0.2$ & $0.92\pm0.04$ & 46 (c) & $-1170\pm30$ \\
\noalign{\smallskip} 
7 & $0.15_{-0.14}^{+0.05}$ & $0.58_{-0.17}^{+0.28}$ & 46 (c) & $-1070\pm40$ \\
\noalign{\smallskip} 
8 & $0.07_{-0.01}^{+0.15}$ & $-0.01_{-0.05}^{+1.82}$ & $790\pm100$ & $-1600\pm800$ \\
\noalign{\smallskip} 
9 & $0.44\pm0.03$ & $-0.65\pm0.06$ & 790 (c) & $-1100\pm140$  \\
\noalign{\smallskip} 
\hline
\end{tabular}
\tablefoot{Microscopic turbulence velocities ($v_{\rm mic}$) followed by (c) are coupled to $v_{\rm mic}$ of another component with the same value in the fit. Statistical uncertainties are quoted at the 68.3~\% confidence level.}
\end{table*}

The hydrogen column density ($N_{\rm H}$) of the warm absorber is dominated by components 1--6 with $N_{\rm H}^{\rm total}\simeq3.8\times10^{26}~{\rm m^{-2}}$, similar to previous results by \citet{net03} with $N_{\rm H}^{\rm total}\sim4\times10^{26}~{\rm m^{-2}}$, which is higher than $N_{\rm H}^{\rm total}\sim2-3\times10^{26}~{\rm m^{-2}}$ found by \citet{kas02} and \citet{kro03}. 

Since the photoionization continuum, abundance table and number of absorption components used by different authors are different, we do not compare the ionization parameters with previous results \citep[e.g.,][]{kas02,net03,kro03}. 

In the present work, the microscopic turbulence velocities of the absorption components are $50~{\rm km~s^{-1}}$ for Components 3--7, $120~{\rm km~s^{-1}}$ for Components 1--2, and $800~{\rm km~s^{-1}}$ for Components 8 and 9, respectively (Table~\ref{tbl:wa_par}). The microscopic turbulence velocity found in the literature also cover a wide range of values, with $\sim100~{\rm km~s^{-1}}$ for \citet{kas00} and \citet{beh03}, $\sim200~{\rm km~s^{-1}}$ for \citet{kro03,net03}, and up to $400-600~{\rm km~s^{-1}}$ for a few ions in Table~3 of \citet{kas02}.

Our best-fit outflow velocities vary for different absorption components, ranging from $-450~{\rm km~s^{-1}}$ to $-1300~{\rm km~s^{-1}}$. A wide range of outflow velocities are also reported in the literature, including $-300$ to $-1000~{\rm km~s^{-1}}$ by \citet{kas02}, $-470$ to $-800~{\rm km~s^{-1}}$ by \citet{beh03}, $-750~{\rm km~s^{-1}}$ by \citet{kro03}, and $-400$ to $-1300~{\rm km~s^{-1}}$ by \citet{net03}. 

We also note that the PION model in the SPEX code takes advantage of recently updated atomic data, which is more accurate and complete than previous models by \citet[K03,][]{kro03} and \citet[N03,][]{net03}. Previous models are able to fit some of the strong absorption lines but insufficient to fit the global spectrum. As pointed out by \citet{net03}, the K03 model provides a better fit to the Fe UTA (unresolved transition array) yet fails to fit the \ion{Si}{X} and \ion{Si}{XI} lines around 6.8 \AA. The N03 model has a better fit over the 5--7 \AA\ wavelength range while the UTA features are poorly fitted. As for the PION model, we refer readers to Section 4.29 of the \href{http://var.sron.nl/SPEX-doc/manualv3.04.00.pdf}{SPEX manual} for a detailed model description, \citet{meh16b} for a comparison with other popular photoionization codes and \citet{mao17b} for a list of characteristic ground and metastable absorption lines.

\subsection{X-ray photoionized emitter}
\label{sct:res_xpe}
\begin{figure*}
\centering
\includegraphics[width=\hsize, trim={0.5cm 0.5cm 0.5cm 0.0cm}, clip]{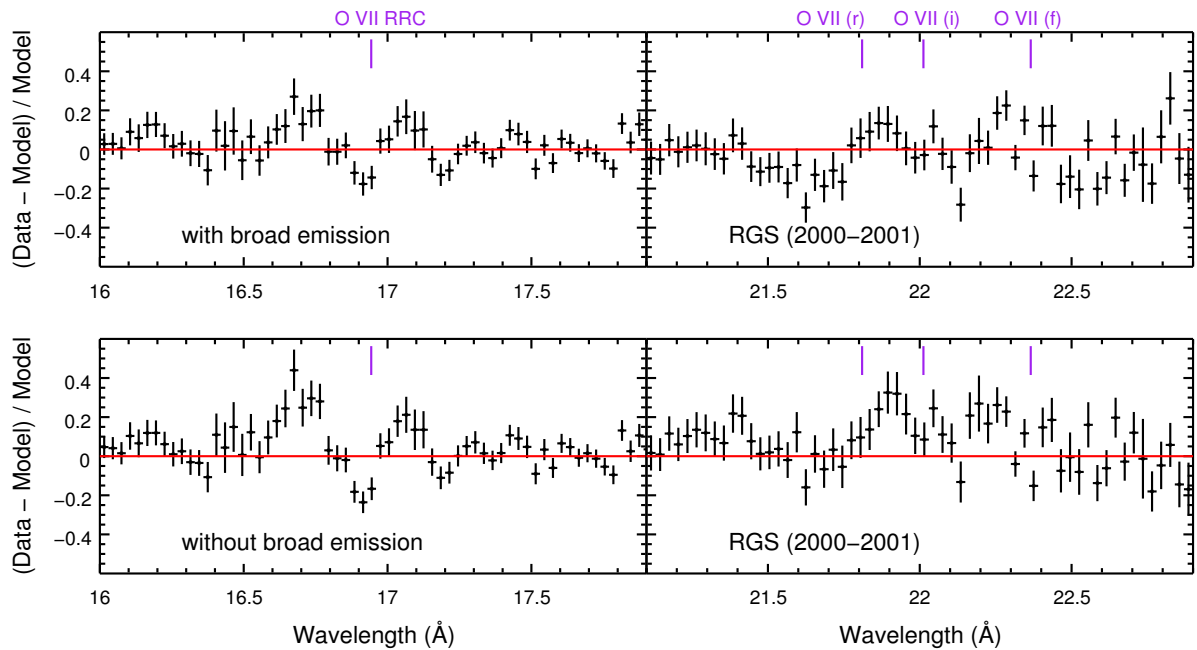}
\caption{Residuals of the best-fit to the time-averaged RGS spectrum (in the observed frame) of NGC\,3783 in 2000--2001 around the \ion{O}{VII} RRC (left panels) and \ion{O}{VII} He-like triplets (right panels). The upper and lower panels show residuals of the best-fit with and without the broad (FWHM$\sim21000~{\rm km~s^{-1}}$) PION emission component (the third component in Table~\ref{tbl:we_par}), respectively.}
\label{fig:dchi_rgs_plot}
\end{figure*}

In our PION modeling of the emission features (lines and RRCs), we find that a broad emission component is required in addition to two narrow emission components. For the time-averaged spectrum, the best-fit with only narrow emission components yields a $C$-stat of 6301 (d.o.f. = 2508), while the best-fit with both broad and narrow emission components yields $\Delta C=-200$ at the price of three degrees of freedom. We tried a few values of the velocity broadening ($7000,~8000,~9000,~{\rm and}~10000~{\rm km~s^{-1}}$), the best-fit $C$-stat. are 6108, 6098, 6092, and 6081, respectively. However, regardless of the presence of the broad emission component, there are still some structures in the residuals of the above two fits (Figure~\ref{fig:dchi_rgs_plot}). Future missions \citep[e.g., \textit{Arcus},][]{smi16} with adequate spectral resolution and significantly larger photon collecting area are essential to verify the presence of such broad emission features in X-rays. 

The broad emission component is obtained via convolving a narrow ($v_{\rm mic}=100~{\rm km~s^{-1}}$) PION component with a velocity broadening model (VGAU in SPEX) with a fixed $\sigma_v$, so that the broadening effect applies to both emission lines and RRCs. Such kind of broadening is probably due to the macroscopic motion of the emitter. Unfortunately, the velocity broadening in X-rays is a difficult parameter to determine. It cannot be too narrow, otherwise there is no significant improvement on the $C$-stat. It cannot be too broad either, otherwise it is unclear whether we are fitting the broad emission line or a part of the complex continuum. Moreover, we are also limited by the calibration uncertainties of the effective area of the instrument, which can easily affect the value of the velocity broadening. At this stage, we tentatively fixed the value of $\sigma_v~(=v_{\rm mac})$ to $9000~{\rm km~s^{-1}}$. This value is within the range of typical gas velocity $3000-10000~{\rm km~s^{-1}}$ \citep{bla90} in the broad line region. For NGC\,3783, the X-ray velocity broadening corresponds to a full width at half maximum (FWHM$~=2.355~\sigma_v$) of about $21000~{\rm km~s^{-1}}$, which is significantly larger than FWHM(H$\beta$) $\sim3000~{\rm km~s^{-1}}$ \citep{onk02} in optical, but close
to the width of the broadest component of Ly$\alpha$ in the UV,
FWHM$\sim18000~{\rm km~s^{-1}}$ \citep{kri18}. 

\begin{figure*}
\centering
\includegraphics[width=\hsize, trim={1.0cm 0.0cm 0.5cm 0.1cm}, clip]{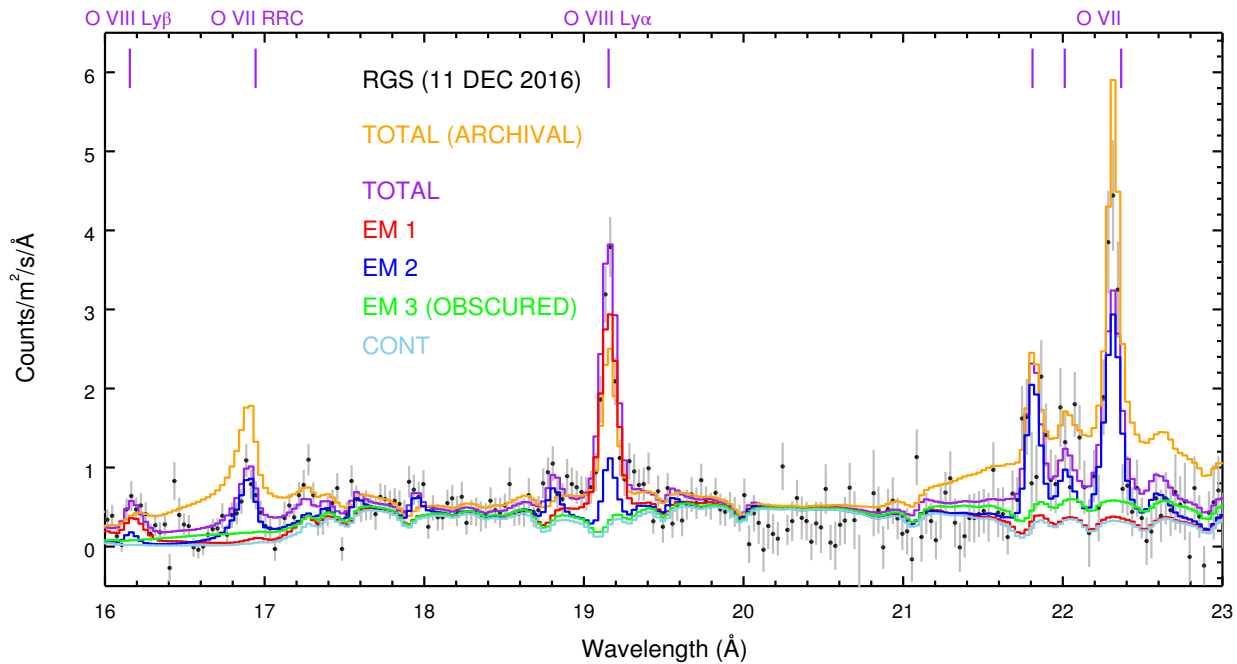}
\caption{Best-fit to the $17-24$~\AA\ RGS spectrum (in the observed frame) of NGC\,3783 from 11 December 2016. The best-fit model with contributions from all emission components is shown in purple. Contributions from individual emission components are shown in different colors, with red and blue for narrow emission features and green for broad emission features. The orange solid line is a calculation using the best-fit parameters obtained from the time-averaged archival spectrum for all three emission components. Due to the obscuration effect, the broad emission component (EM 3) appears to be weaker in December 2016. }
\label{fig:spec_161211_rgs_zoom_plot}
\end{figure*}

\begin{figure}
\centering
\includegraphics[width=\hsize, trim={0.3cm 0.0cm 0.5cm 0.1cm}, clip]{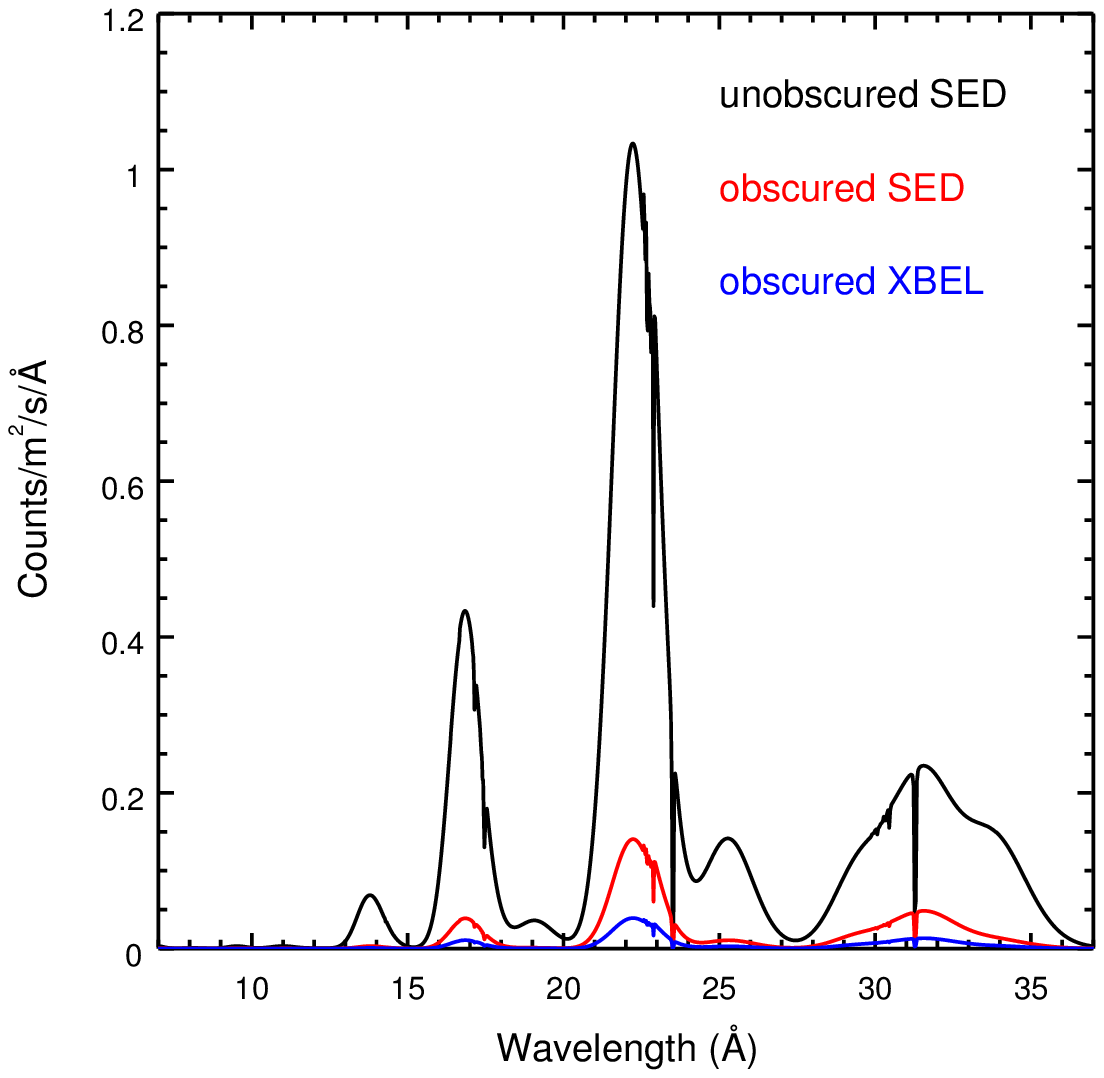}
\caption{Photoionization calculations of the X-ray broad emission lines (XBEL). In all three cases below, line profiles are corrected for the redshift and Galactic absorption of NGC\,3783. The black solid curve corresponds to a case where the XBEL is directly exposed to the intrinsic SED of 11 December 2016. The red curve is for a case where the XBEL is exposed to the obscured SED of 11 December 2016. The blue curve is for a case where the XBEL is photoionized by the intrinsic SED of 11 December 2016 and further screened by the obscurer.}
\label{fig:xbel_plot}
\end{figure}

The broad and narrow emission model components derived from the 2000--2001 data do not match the  data observed in December 2016 (the orange line in Figure~\ref{fig:spec_161211_rgs_zoom_plot}). Narrow emission features might have varied slightly over the 15-year timescale (Section~\ref{sct:res_var}). The broad emission component seems to be much weaker in December 2016 (Figure~\ref{fig:spec_161211_rgs_zoom_plot}). To explore possible explanations, we performed three sets of photoionization calculation of the X-ray broad emission lines, assuming the 2000--2001 parameters for X-ray broad emission lines. The first case is the simplest one where the XBEL is directly exposed to the intrinsic SED (the black curve in Figure~\ref{fig:xbel_plot}). In the presence of the obscurer, we need to consider two more cases. The XBEL might be exposed to the obscured SED (the red curve in Figure~\ref{fig:xbel_plot}), or the XBEL is photoionized by the intrinsic SED but further screened by the obscurer (the blue curve in Figure~\ref{fig:xbel_plot}). Among all three cases, the last one yields the weakest broad emission profiles. The second case yields a slight increase of $C$-stat ($\Delta C=+5$) compared to the last one. Therefore, the apparent weakening observed in 2016 can be accounted for by applying the obscuration to the broad emission component (the purple line in Figure~\ref{fig:spec_161211_rgs_zoom_plot}). But we cannot strictly rule out the possibility that a fraction of the obscurer is co-existing or even closer to the black hole with respect to the X-ray broad-line region. The parameters for the obscurer used here are the same as given in Table~1 of \citet{meh17}. The observed flux of the broad emission component around the \ion{O}{VII} He-like triplets (21.0--23.6~\AA) decreases from $23\times10^{-17}~{\rm W~m^{-2}}$ (2000--2001, i.e., ``unobscured SED" in Figure~\ref{fig:xbel_plot}) to $4\times10^{-17}~{\rm W~m^{-2}}$ (``obscured SED"), and $1\times10^{-17}~{\rm W~m^{-2}}$ (December 2016, i.e., ``obscured XBEL"). The above interpretation supports the picture suggested by \citet{meh17} that the obscurer is currently at the outer broad-line region of the AGN. Admittedly, if our assumption is not valid that the X-ray broad emission component varies significantly (column density, ionization parameter and broadening), we can no longer break the degeneracy shown in Figure~\ref{fig:xbel_plot}. 

\begin{figure*}
\centering
\includegraphics[width=\hsize, trim={0.5cm 1.0cm 0.5cm 0.1cm}, clip]{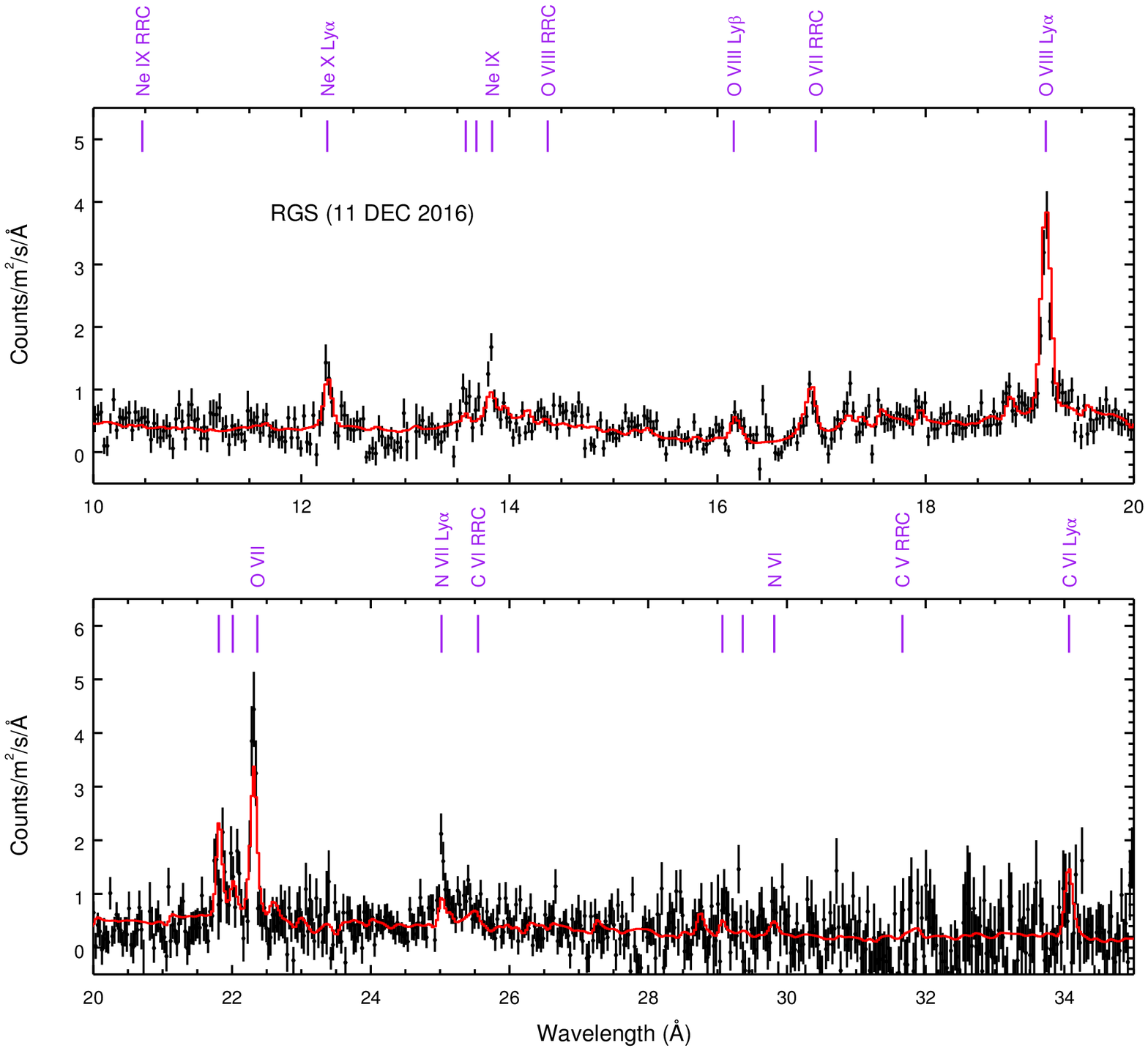}
\caption{The best-fit to the RGS spectrum (in the observed frame) of NGC\,3783 on 11 DEC 2016. Most prominent emission features are labeled. }
\label{fig:spec_161211_rgs_plot}
\end{figure*}

The best-fit to the RGS spectrum on 11 December 2016 is shown in Figure~\ref{fig:spec_161211_rgs_plot}, which is similar to that on 21 December 2016. The best-fit parameters of the X-ray photoionized emitter in 2000--2001 and December 2016 are listed in Table~\ref{tbl:we_par}. While the emission measures (E.M.$=n_e~n_{\rm H}~4\pi~C_{\rm em}~r^2~N_{\rm H}/n_{\rm H}$) of component 1 are consistent (at $1\sigma$ confidence level) between epochs, the emission covering factor ($C_{\rm em}$), thus the emission measure of component 2 is an order of magnitude higher. If we fix $C_{\rm em}(\rm EM~2)=0.6$ when fitting the 11 December 2016 spectrum, the best-fit $C$-stat is 2406, which is a worse fit ($\Delta C\sim+70$) when compared to the best-fit $C$-stat in Table~\ref{tbl:sed_par}. Meanwhile, the ionization parameter of component 2 is a factor of two smaller in 2016, accordingly the line emissivity is an order of magnitude lower in 2016. This might suggest an increase of the physical size of the emitter over 15 years. 

\begin{table*}
\caption{Best-fit results of the X-ray photoionized emitter in the time-averaged spectra of NGC\,3783. The emission measure (E.M.) is calculated based on the best-fit parameters. }
\label{tbl:we_par}
\centering
\begin{tabular}{cccccccccccccccc}
\hline\hline
\noalign{\smallskip}
Comp. & $N_{\rm H}$ & $\log_{10}(\xi)$ & $v_{\rm mic}$ & $C_{\rm em}$ & E.M. \\
& $10^{26}~{\rm m^{-2}}$ & $10^{-9}~{\rm W~m}$ & ${\rm km~s^{-1}}$ & \% & $10^{70}~{{\rm m}^{-3}}$ \\
\noalign{\smallskip} 
\hline
\noalign{\smallskip} 
\multicolumn{6}{c}{2000--2001} \\
\noalign{\smallskip} 
1 & $60\pm37$ & $2.60\pm0.07$ & $600\pm100$ & $0.29\pm0.12$ & $0.5_{-0.4}^{+0.8}$ \\
\noalign{\smallskip}
2 & $5.2_{-2.3}^{+5.4}$ & $1.35\pm0.05$ & $140_{-100}^{+50}$ & $0.6_{-0.2}^{+0.4}$ & $1.5_{-1.0}^{+4.1}$ \\
\noalign{\smallskip}
3 & $28_{-5}^{+72}$ & $0.82\pm0.02$ & 100 (f) & $0.3_{-0.2}^{+0.1}$ & $13_{-10}^{+53}$ \\
\noalign{\smallskip} 
\hline
\noalign{\smallskip} 
\multicolumn{6}{c}{11 DEC 2016} \\
\noalign{\smallskip} 
1 & $25\pm6$ & $2.58\pm0.05$ & $590\pm90$ & $0.98\pm0.13$ & $1.0_{-0.4}^{+0.6}$ \\
\noalign{\smallskip}
2 & $3.0\pm0.7$ & $1.03\pm0.05$ & $350\pm70$ & $7.0\pm1.3$ & $30_{-13}^{+19}$ \\
\noalign{\smallskip}
3 & 28 (f) & 1.00 (f) & 100 (f) & 0.3 (f) & 13 (f) \\
\noalign{\smallskip} 
\hline
\noalign{\smallskip} 
\multicolumn{6}{c}{21 DEC 2016} \\
\noalign{\smallskip} 
1 & $35\pm17$ & $2.58\pm0.05$ & $460\pm130$ & $0.7\pm0.3$ & $1.6_{-0.9}^{+1.6}$ \\
\noalign{\smallskip}
2 & $2.3\pm1.0$ & $1.03\pm0.05$ & $270\pm110$ & $7.3\pm2.4$ & $28_{-19}^{+32}$ \\
\noalign{\smallskip}
3 & 28 (f) & 1.06 (f) & 100 (f) & 0.3 (f) & 13 (f) \\
\noalign{\smallskip} 
\hline
\end{tabular}
\tablefoot{Parameters followed by (f) are fixed in the fit. The 2000--2013 MEG ($6-17$~\AA) data and the 2000--2001 RGS ($7-37$~\AA) data are fitted simultaneously (Section~\ref{sct:spec_data}), but constraints are obtained mainly from the 2000--2001 RGS data since it covers all the emission features from Ne, O, N, and C. Statistical uncertainties are quoted at the 68.3~\% confidence level.}
\end{table*}

Different values of the best-fit parameters of the narrow emission components in different epochs (Table~\ref{tbl:we_par}) need to be interpreted with caution. Due to lack of information, we assume the photoionization continuum to be the 2000--2001 AGN SED for all spectra in 2000-2001 and December 2016. Under this assumption, the ionization parameter $\xi = L/(n_{\rm H}~r^2)$ should be the same, because the number density ($n_{\rm H}$) and distance ($r$) of the emission region are not expected to vary dramatically. The column density ($N_{\rm H}$), microscopic turbulence velocity ($v_{\rm mic}$) and emission covering fraction ($C_{\rm em}$) might not vary significantly as well. Deviation from the time-average photoionization continuum can be partly due to long-term variation in the intrinsic SED components and/or the obscuration events, which might be more frequent than previously thought for NGC\,3783. Based on the hardness ratio of all \textit{Swift}/XRT spectra in 2008--2017, obscuration events might occur about half of the time \citep{kaa18}, which will affect the photoionization modelling of the X-ray narrow emission features.

Additionally, by comparing the best-fit results of the absorption (Table~\ref{tbl:wa_par}) and emission (Table~\ref{tbl:we_par}) components, we find that emission component 1 and absorption component 3 share similar ionization parameter, yet the hydrogen column density, turbulence velocity and outflow velocity are significantly different. The ionization parameter of emission component 2 has no counterpart in absorption components. This is similar to what we find in another Seyfert 1 galaxy NGC\,5548 \citep{mao18a}. It is possible that the X-ray emission and absorption components are not related, although we need distance (thus density) measurement on these components to verify this deduction.

\subsection{Variability of the X-ray emission features}
\label{sct:res_var}
We check the variability of the most prominent narrow emission lines, the \ion{O}{VIII} Ly$\alpha$ line and \ion{O}{VII} He-like triplets, in the RGS spectra in 2000--2001 and December 2016. A phenomenological local fit is used in this exercise so that we are not confused by the unknown photoionization continuum effect on the photoionization modeling. 

\begin{figure}
\centering
\includegraphics[width=\hsize, trim={0.0cm 0.0cm 0.5cm 0.1cm}, clip]{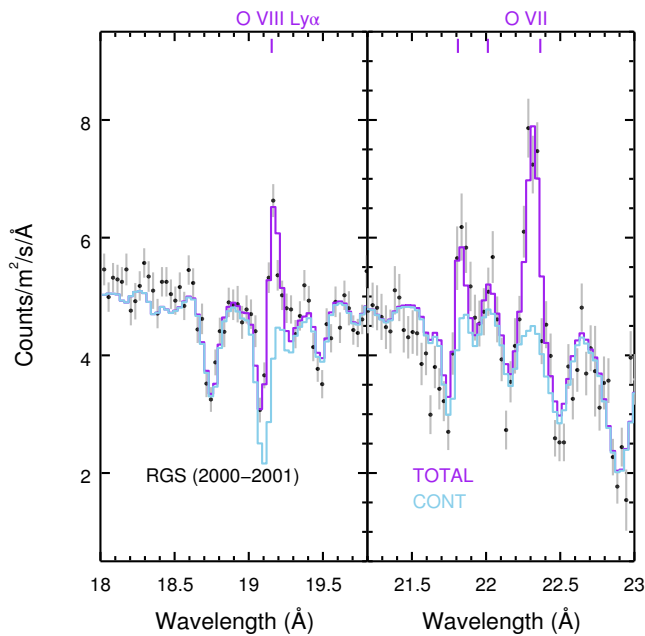}
\caption{Local fit to the observed \ion{O}{VIII} Ly${\alpha}$ and \ion{O}{VII} He-like triplets in the 2000-2001 time-averaged RGS spectrum of NGC\,3783. The continuum model (CONT, light blue) is fixed to the best-fit global continuum, accounting for the foreground Galactic absorption, the warm absorber and including the broad emission features in addition. }
\label{fig:spec_00to13_rgs_delta_plot}
\end{figure}

\begin{figure}
\centering
\includegraphics[width=\hsize, trim={0.0cm 0.0cm 0.5cm 0.1cm}, clip]{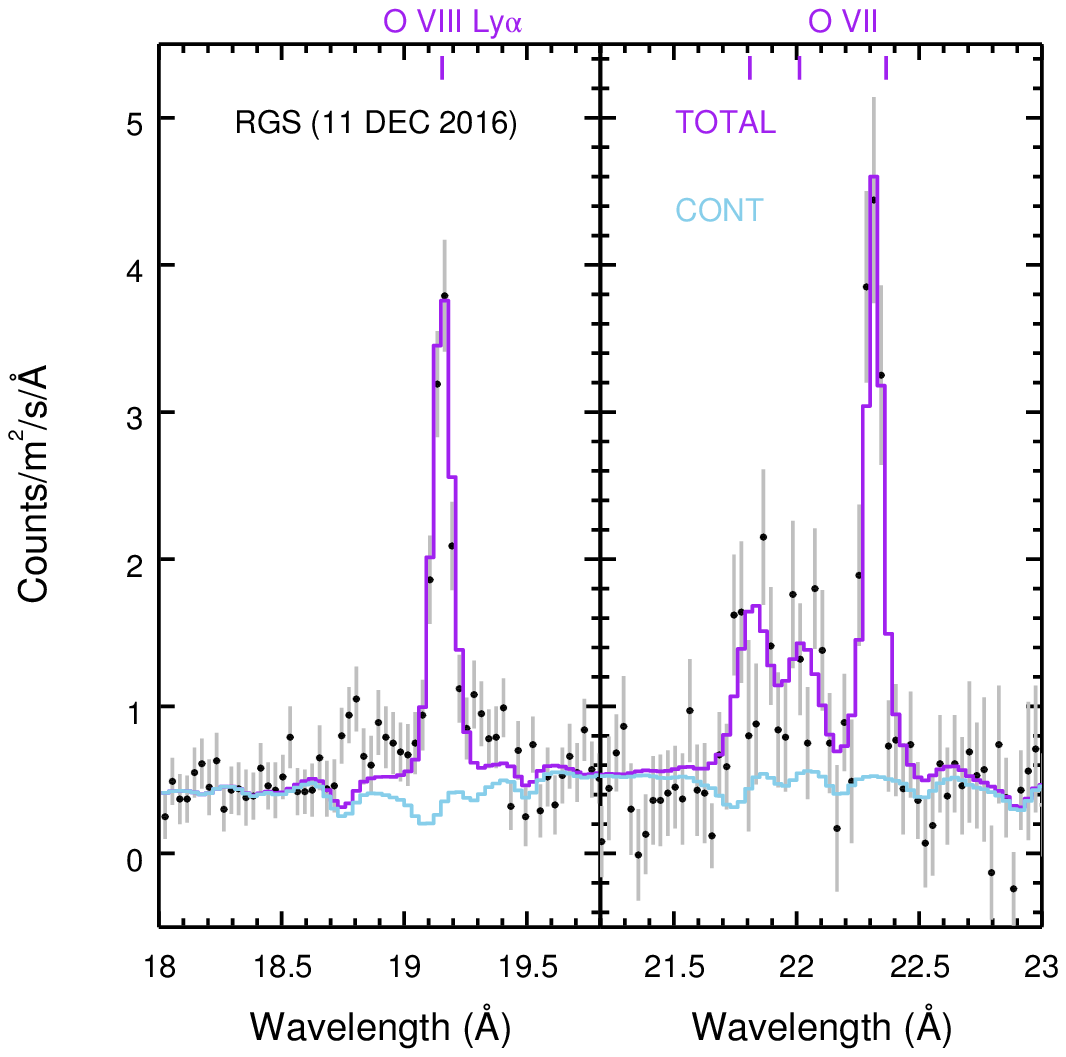}
\caption{Local fit to the observed \ion{O}{VIII} Ly${\alpha}$ (left panel) and \ion{O}{VII} He-like triplets (right panel) in the RGS spectrum of NGC\,3783 from 11 December 2016. The continuum model (CONT, light blue) is fixed to the best-fit global continuum, accounting for the foreground Galactic absorption, the obscurer, the warm absorber and including the obscured broad emission features in addition. }
\label{fig:spec_161211_rgs_delta_plot}
\end{figure}

We fix the continuum model to the best-fit global continuum accounting for the foreground Galactic absorption, the obscurer, the warm absorber and including the broad emission features in addition. Subsequently, the narrow emission lines are accounted for with Gaussian line profiles. The normalization and velocity broadening of the Gaussian profiles are free to vary, except that we limit the velocity broadening to be no larger than $2000~{\rm km~s^{-1}}$ and couple the broadening of the resonance and intercombination lines of the \ion{O}{VII} He-like triplets. The line luminosity and velocity broadening are listed in Table~\ref{tbl:delta_par}. The best-fit results of 2000--2001 and 11 December 2016 are plotted in Figs.~\ref{fig:spec_00to13_rgs_delta_plot} and \ref{fig:spec_161211_rgs_delta_plot}. 

\begin{table*}
\caption{Best-fit results of the oxygen emission lines from RGS data in 2000--2001 and December 2016.}
\label{tbl:delta_par}
\centering
\begin{tabular}{cccccccccccccccc}
\hline\hline
\noalign{\smallskip}
Ion & Line & \multicolumn{2}{c}{2000--2001} & \multicolumn{2}{c}{11 DEC 2016} & \multicolumn{2}{c}{21 DEC 2016} \\
& $\lambda$ ($\AA$) & $L~(10^{32}~{\rm W})$ & $\sigma_v~({\rm km~s^{-1}})$ & $L~(10^{32}~{\rm W})$ & $\sigma_v~({\rm km~s^{-1}})$ & $L~(10^{32}~{\rm W})$ & $\sigma_v~({\rm km~s^{-1}})$ \\
\noalign{\smallskip} 
\hline
\noalign{\smallskip} 
\ion{O}{VIII} Ly$\alpha$ & 18.97 & $13.4\pm0.9$ & $360\pm130$ & $17.8\pm1.1$ & $370\pm100$ & $19.8\pm1.1$ & $520\pm130$ \\
\noalign{\smallskip} 
\ion{O}{VII} (r) & 21.60 & $6.6\pm1.4$ & $<280$ & $9.2\pm1.8$ & $740\pm190$ & $8.4\pm2.1$ & $<500$ \\
\noalign{\smallskip} 
\ion{O}{VII} (i) & 21.81 & $1.3\pm1.3$ & $<280$ (c) & $6.0\pm1.6$ & $740$ (c) & $5.5\pm1.0$ & $<500$ (c) \\
\noalign{\smallskip} 
\ion{O}{VII} (f) & 22.10 & $18.5\pm1.7$ & $310\pm100$ & $17.6\pm1.9$ & $<180$ & $17.3\pm3.1$ & $<300$ \\
\noalign{\smallskip} 
\hline
\end{tabular}
\tablefoot{Velocity broadening ($\sigma_v$) of the resonance and intercombination lines of \ion{O}{VII} are coupled (c) in the fit. Statistical uncertainties are quoted at the 68.3~\% confidence level.}
\end{table*}

These emission lines are consistent with each other at a $1\sigma$ confidence level between the two observations in December 2016. When compared to 2000--2001, the \ion{O}{VII} resonance and forbidden lines remain constant in luminosity at the $1\sigma$ confidence level, while the \ion{O}{VIII} Ly$\alpha$ line and intercombination line of He-like \ion{O}{VII} are marginally brighter in December 2016 at the $2\sigma$ confidence level. In addition, the resonance and intercombination lines of \ion{O}{VII} appear to be broader in December 2016, although the uncertainty is large. Note that the line luminosity depends on both the emission measure and line emissivity. While the emission measure of component 2 (Section~\ref{sct:res_xpe}) increases in 2016, the line emissivity decreases, leading to the comparable line luminosity between epochs.

We should point out that the residuals on both sides of the \ion{O}{VIII} Ly$\alpha$ line in Figure~\ref{fig:spec_161211_rgs_delta_plot} are accounted for in the full plasma model fit shown in Figure~\ref{fig:spec_161211_rgs_zoom_plot}. A full plasma model includes many dielectric satellite lines, which are even weaker than the \ion{O}{VII} He$\beta$ line (at $18.63~\AA$).

Since the variability of the X-ray broad emission features cannot be checked directly, we turn to the very broad (FWHM$\gtrsim10^4~{\rm km~s^{-1}}$) emission lines in the high-resolution UV spectra with HST in 2000--2001 and December 2016 \citep{kri18}. No significant variation is found in the line flux of Ly$\alpha$, \ion{Si}{IV}, \ion{C}{IV} and \ion{He}{II} between the two epochs. Hence, our previous assumption of a non-varying X-ray broad emission component is reasonable. 

\subsection{Summary}
\label{sct:sum}
\begin{figure}
\centering
\includegraphics[width=\hsize, trim={0.0cm, 0.0cm, 0.0cm, 0.0cm}, clip]{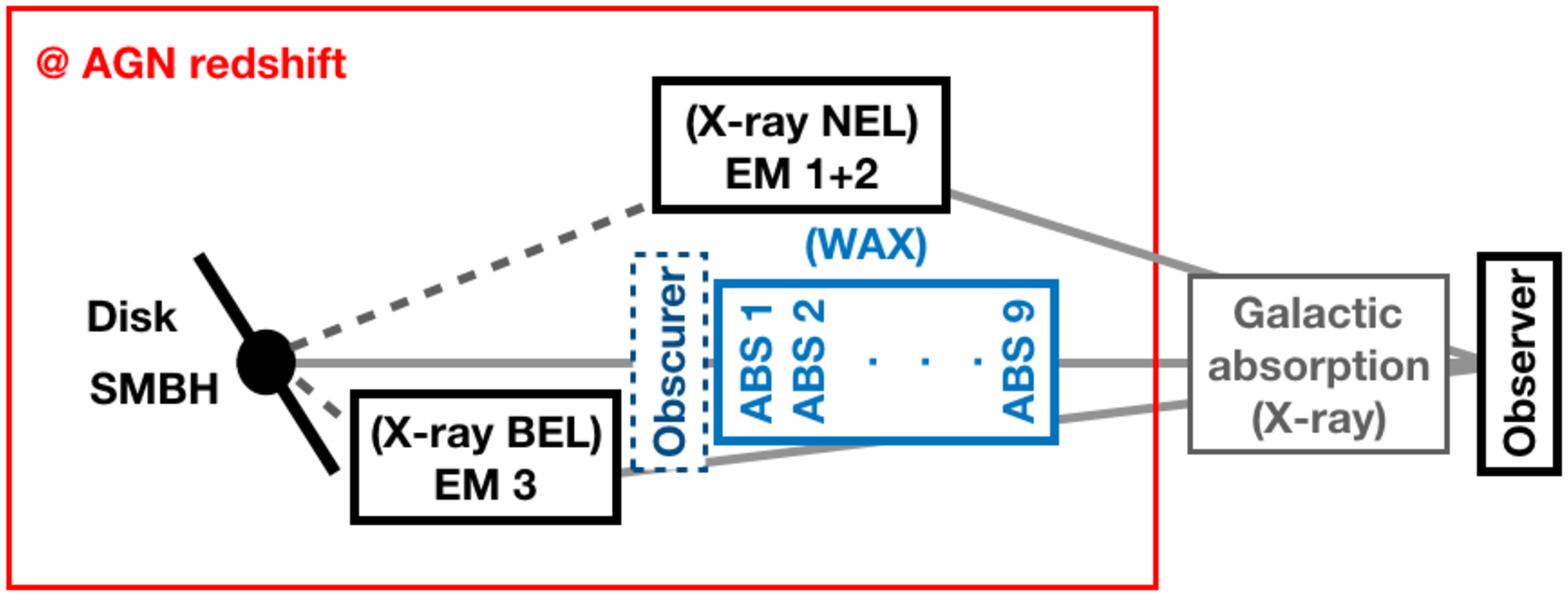}
\caption{Simplified view of the relations among spectral components observed in NGC\,3783 in X-rays. Three different light paths are shown here. The obscurer (dashed box) is not always present along the line of sight. Dashed lines indicate that these photons are not directly observable. WAX is short for the warm absorber in X-rays. BEL and NEL refer to the broad and narrow emission lines. EM and ABS are short for emission and absorption, respectively.}
\label{fig:com_rel}
\end{figure}

We analyzed the X-ray spectra of the Seyfert 1 galaxy NGC\,3783 using both archival data in 2000--2013 and newly obtained data in December 2016. The intrinsic SED of the AGN, the obscuration, absorption, emission, and extinction effects are fitted simultaneously. In Figure~\ref{fig:com_rel} we show a sketch of the relations among spectral components resulting from our modelling. 

Along the line of sight toward the nucleus, prominent absorption line features due to the warm absorber were clearly visible in the 2000--2013 X-ray spectra \citep[e.g.,][]{net03,kro03,sco14}. Nine X-ray absorption components with different ionization parameters and kinematics are required in our photoionization modeling. On 11 and 21 December 2016, obscuration was seen in NGC\,3783 \citep{meh17}. The obscurer produces the heavy absorption of the soft X-ray continuum, as well as broad and blueshifted absorption lines in the UV spectra. It is very likely that there are more obscuration events in NGC\,3783 \citep{mar14,kaa18}.

The X-ray narrow emission components reprocess photons from the nucleus that are not directly observed (dashed lines in Figure~\ref{fig:com_rel}). Prominent emission features like the \ion{O}{VIII} Ly$\alpha$ line  and the He-like triplets of \ion{O}{VII} are visible whether the soft X-ray continuum is obscured or not. Weak emission features like the \ion{O}{VII} RRC are detectable only when the continuum is obscured. Our photoionization modeling requires two components for the X-ray narrow emission features in NGC\,3783. The two narrow emission components have different emission measure, ionization parameter, turbulence velocity, and emission covering factor. This is similar to our previous analysis for NGC\,4151 \citep[Seyfert 1.5,][]{arm07}, Mrk\,335 \citep[Seyfert 1,][]{lon08}, NGC\,4051 \citep[Seyfert 1,][]{nuc10}, NGC\,424 \citep[Seyfert 2,][]{mar11}, and NGC\,5548 \citep[Seyfert 1.5,][]{mao18a}. On the other hand, \citet{gua09} and \citet{whe15} model the X-ray narrow emission features in NGC\,1365 (Seyfert 1.8) and NGC\,5548 (Seyfert 1.5), respectively, with only one photoionized component. 

Our photoionization modeling (Table~\ref{tbl:we_par}) finds that the ionization parameter of the X-ray narrow emission component 1 remains constant throughout 2000--2016, indicating that either this component has a distance of at least a few light-years or its density is very low. On the other hand, for the X-ray narrow emission component 2, the ionization parameters are consistent between the two observations in December 2016 yet about a factor of two lower when compared to 2000--2001. If the variation is due to the change in the ionizing continuum, this suggests that component 2 is closer than component 1, but its distance should be larger than 10 light-days. 

The X-ray broad emission component also reprocesses photons from the nucleus. In this work, we demonstrate that, when the obscurer is present in NGC\,3783, it screens photons emitted from the X-ray broad emission component, which leads to the apparent weakening of the X-ray broad emission features in December 2016 (Section~\ref{sct:res_xpe}). The above interpretation supports the geometry proposed by \citet{meh17} that the obscurer is located farther away than the X-ray broad emission component, which is a few light days from the nucleus \citep{pet04}. 
Our photoionization modeling of the observed X-ray broad emission features is an ad hoc interpretation. There is evidence that the optical, UV, and X-ray broad emission lines originate from the same photoionized plasma \citep[e.g.,][]{cos16}. Thus, the broad-line region  has a range of densities, ionization parameters, and kinematics. A dedicated analysis with the ``locally optimally emitting cloud" approach \citep[e.g.,][]{cos07} or even sophisticated dynamic models \citep[e.g.,][]{pan11} is required to interpret the optical to X-ray broad emission lines in NGC\,3783. 

For simplicity, we assume that the X-ray broad and narrow emission components are not further absorbed by the warm absorber (Section~\ref{sct:spec_model}), which might not be true. That is to say, the unabsorbed luminosity of the X-ray broad and narrow emission lines obtained here is only a lower limit. To properly account for the screening effects by the warm absorber, we need to know which warm absorber components are more distant than the X-ray emission regions, and what are the covering factors for these warm absorber components with respect to the X-ray emission regions. 

\section{Conclusions}
\label{sct:con}
We focus on the photoionized emission features in the high-resolution X-ray spectra of NGC\,3783 obtained in December 2016 when the soft X-ray continuum was heavily obscured. We also analyze the archival time-averaged high-resolution X-ray spectrum in 2000--2001 to compare the photoionized emission features and study the warm absorber. The main results are summarized as follows.

\begin{enumerate}
    \item Nine photoionization components with different ionization parameters and kinematics are required for the warm absorber. 
    \item Two photoionization components are required for the X-ray narrow emission features, which are weakly varying over the past 15 years. 
    \item The presence of a X-ray broad emission component significantly improves the fit to the time-averaged spectrum in 2000--2001.
    \item The X-ray broad emission features are much weaker in December 2016. This apparent weakening can be explained by the obscuration effect on the X-ray broad emission component.
\end{enumerate}

\begin{acknowledgements}
      We thank the referee for useful comments and suggestions. This work is based on observations obtained with \textit{XMM-Newton}, an ESA science mission with instruments and contributions directly funded by ESA Member States and the USA (NASA). This research has made use of data obtained with the \textit{NuSTAR} mission, a project led by the California Institute of Technology (Caltech), managed by the Jet Propulsion Laboratory (JPL) and funded by NASA. This work made use of data supplied by the UK \textit{Swift} Science Data Centre at the University of Leicester. SRON is supported financially by NWO, the Netherlands Organization for Scientific Research. This work was supported by NASA through a grant for HST program number 14481 from the Space Telescope Science Institute, which is operated by the Association of universities for Research in Astronomy, Incorporated, under NASA contract NAS5-26555. The research at the Technion is supported by the I-CORE program of the Planning and Budgeting Committee (grant number 1937/12). EC is partially supported by the NWO-Vidi grant number 633.042.525. CP acknowledges funding through ERC Advanced Grant number 340442. EB is grateful for funding from the European Union's Horizon 2020 research and innovation programme under the Marie Sklodowska-Curie grant agreement no. 655324. SB acknowledges financial support from ASI under grant ASI-INAF I/037/12/0, and from the agreement ASI-INAF n. 2017-14-H.O. GP acknowledges support from the Bundesministerium f\"{u}r Wirtschaft und Technologie/Deutsches Zentrum f\"{u}r Luft- und Raumfahrt (BMWI/DLR, FKZ 50 OR 1408) and the Max Planck Society. BDM acknowledges support by the Polish National Science Center grant Polonez 2016/21/P/ST9/04025. POP aknowledges support from the CNES and French PNHE. LDG acknowledges support from the Swiss National Science Foundation. 
\end{acknowledgements}


\begin{appendix}
\section{Component relation in fitting the spectra using SPEX}
\label{sct:com_rel}
The emission and absorption components of our model for the X-ray spectrum of NGC 3783 have a complex spatial relationship that is important for how we construct our model in SPEX. Figure~\ref{fig:com_rel} gives a visual representation of our model. The intrinsic SED (model component 1 in Section~\ref{sct:spec_model}) originates in the accretion disk surrounding the supermassive black hole (SMBH). This is the innermost component of the model. This innermost emission region is absorbed by both the obscurer, when it is present (model component 2), and the X-ray warm absorber (WAX). The next emission layer is the X-ray BEL (part of model component 4). This is also absorbed by the obscurer when it is present. The outermost emission component is the X-ray NEL (part of model component 4), which may be co-spatial with the WAX, but we presume it is not absorbed by the WAX. The final component in our model is Galactic absorption (model component 7), which affects all other components originating in NGC\,3783.

The intrinsic SED of NGC 3783 consists of a Comptonized disk component (COMT in SPEX), the power-law component (POW), and a neutral reflection component (REFL). These are additive components. An exponential cut-off is applied to the high- and low- energy end of the power-law component. This is realized via the ETAU model in SPEX (see also Chapter 7 of the \href{http://var.sron.nl/SPEX-doc/cookbookv3.00.00.pdf}{SPEX cookbook}), which provides a simple transmission $T(E)=\exp(-\tau_0~E^a)$, where $\tau_0$ is the optical depth at 1~keV. For the high-energy cut-off, we set $a=1$ and $\tau_0 = 1/340$, which corresponds to a cut-off energy at $340$~keV \citep{dro02}. For the low-energy cut-off, we set $a=-1$ and $\tau_0 = T_{\rm seed}$, which corresponds to a cut-off at the seed photon temperature ($T_{\rm seed}$) of the Comptonized disk component. 

The warm absorber is modeled with nine PION absorption components, which are multiplicative components. The most highly ionized component (index 1) is assumed to be the closest to the nucleus, while the least ionized component (index 9) is assumed to be the furthest. In the absence of the obscurer, the first PION absorption component is directly exposed to the AGN SED. Ionizing photons received by the second PION absorption is the AGN SED screened by the first PION absorption component. The rest is done in the same manner. In the presence of the obscurer, photons from the nucleus are first screened by the obscurer then modified by PION absorption components. 

The PION emission components (for the X-ray photoionized emitter) are both multiplicative and additive. This is because they reprocess ionizing photons which are not directly observed (dashed lines in Figure~\ref{fig:com_rel}) and emit photons which are directly observed. Accordingly, the ionizing continuum is first multiplied by the PION emission components (as multiplicative components) then followed by a ETAU component with $\tau_0=1000$ and $a=0$ so that this ionizing continuum is not present in the spectrum. As additive components, the PION emission components are multiplied by the redshift of the AGN and the Galactic absorption in X-rays. 

We provide a rough estimate of the distance of the X-ray photoionized emitter in the following. To estimate the distance of the narrow and broad line gas, we need the mass of the supermassive black hole, the geometry factor, broadening of the line due to Keplerian motion \citep[e.g., Eq. 1 of][]{pet04}. The geometry factor is usually assumed to be 5.5 \citep{pet04}. But the broadening of the line due to Keplerian motion is unknown in our case. We do have measurement on the line broadening but, for the narrow emission line components, we attribute the broadening to microscopic turbulence. Note that turbulence does not only affect the broadening but also the strength of the line. We attempted to add a macroscopic motion to the narrow line gas, but we cannot obtain a meaningful constrain ($v_{\rm mac}\ll v_{\rm mic}$). That is to say, we only have the upper limit on the line broadening due to Keplerian motion, which yields the lower limit of the distance (> 1 pc) of the narrow-line gas.

For the broad-line gas, the X-ray FWHM is broader than that of the optical, indicating that the X-ray emitting gas is closer to the black hole than the optical emitting gas. The distance of the optical BLR ranges from 1.4 light-day to 10.2 light-day \citep{pet04}. Thus, we have a reasonable upper limit of the distance (<1.4 light-day) of the X-ray broad-line gas. 

The distance of the obscurer is about 10 light-day, according to \citet{meh17}. Thus, the obscurer is farther away than the X-ray BLR but closer than the X-ray NLR.

\end{appendix}

\end{document}